\documentclass{article}

\usepackage{arxiv}

\usepackage[utf8]{inputenc} 
\usepackage[T1]{fontenc}    
\usepackage{hyperref}       
\usepackage{url}            
\usepackage{booktabs}       
\usepackage{amsfonts}       
\usepackage{nicefrac}       
\usepackage{microtype}      
\usepackage{lipsum}
\usepackage{graphicx}

\usepackage{siunitx}                 

\graphicspath{ {./images/} }

\title{Tripartite quantum correlations obtained by post-selection from twin beams}

\author{
  Pavel Pavl\'{\i}\v{c}ek \\
  Institute of Physics of the Czech Academy of Sciences,\\
  Joint Laboratory of Optics of Palacky University and Institute of Physics of CAS,\\
  17. listopadu 50a, 779 00 Olomouc, Czech Republic \\
  \texttt{Pavel.Pavlicek@upol.cz} \\
   \And
 Jan Pe\v{r}ina Jr. \\
  Joint Laboratory of Optics of Palacký University and Institute of Physics of CAS,\\
  Faculty of Science, Palacký University,\\
  17. listopadu 12, 779 00 Olomouc, Czech Republic \\
  \texttt{Jan.Perina.Jr@upol.cz}\\
  \And
 V\'{a}clav Mich\'{a}lek \\
  Institute of Physics of the Czech Academy of Sciences,\\
  Joint Laboratory of Optics of Palacky University and Institute of Physics of CAS,\\
  17. listopadu 50a, 779 00 Olomouc, Czech Republic \\
  \texttt{Vaclav.Michalek@upol.cz} \\
 \And
  Radek Machulka \\
  Institute of Physics of the Czech Academy of Sciences,\\
  Joint Laboratory of Optics of Palacky University and Institute of Physics of CAS,\\
  17. listopadu 50a, 779 00 Olomouc, Czech Republic \\
  \texttt{Radek.Machulka@upol.cz} \\
	\And
	Ond\v{r}ej Haderka \\
  Joint Laboratory of Optics of Palacký University and Institute of Physics of CAS,\\
  Faculty of Science, Palacký University,\\
  17. listopadu 12, 779 00 Olomouc, Czech Republic \\
  \texttt{Ondrej.Haderka@upol.cz}\\
}

\begin{document}
\maketitle
\begin{abstract}
Spatially-resolved photon counting of a twin beam performed by an
iCCD camera allows for versatile tailoring the properties of the
beams formed by parts of the original twin beam. Dividing the
idler beam of the twin beam into three equally-intense parts and
post-selecting by detecting a given number of photocounts in the
whole signal beam we arrive at the idler fields exhibiting high
degrees of nonclassicality and being endowed with tripartite
quantum correlations. Nonclassicality is analyzed with the help of
suitable nonclassicality witnesses and their corresponding
nonclassicality depths. Suitable parameters are introduced to
quantify quantum correlations. These parameters are analyzed as
they depend on the field intensity. The experimental photocount
histograms are reconstructed by the maximum-likelihood approach
and the obtained photon-number distributions are compared with a
suitable model in which the original twin beam is approximated by
an appropriate multi-mode Gaussian field and undergoes the
corresponding beams' transformations. 
\end{abstract}

\keywords{Spontaneous parametric down-conversion; Twin-beam;
Three-mode correlations, Photon-number anti-correlations,
Nonclassicality}

\section{Introduction}
A twin beam (TWB) is formed by a pair of correlated light beams
generated in the process of spontaneous parametric down-conversion
or in the process of nonlinear optical
amplification.~\cite{Boyd2003}. The constituting beams are
historically referred to as the signal and idler beams. Twin beams
are endowed with nonclassical properties.~\cite{Mandel1995} They
exhibit the entanglement between the signal and idler beams. The
entanglement originates in the correlation in the number of
photons, entanglement also arises in the beams' polarization,
spectral correlations, and tight spatial
correlations.~\cite{Genovese2016, Jedrkiewicz2004, Haderka2005a,
Bondani2007, Blanchet2008}

Twin beams are suitable for generating the states with the
sub-Poissonian distribution using the post-selection process.
Post-selection is performed based on the detection of a given
number of photons in one of the two
beams.~\cite{Bondani2007,Rarity1997,Laurat2003,Zou2006,PerinaJr2013b,Lamperti2014,Iskhakov2016,Harder2016}
The states with sub-Poissonian distribution are useful for quantum
measurements as well as quantum imaging with the errors being
below the standard quantum limit.~\cite{Jakeman1986,
SabinesChesterkind2019}

The post-selection process can be applied to more complex fields
composed of several TWBs in different configurations. It is shown
in Ref.~\cite{PerinaJr2021,PerinaJr2021a} that a two-beam field
with anti-correlations in photon-number fluctuations and marginal
sub-Poissonian photon-number distributions is left in the idler
beams of two TWBs sharing their signal beams after post-selection
that is based on the measurement of a given number of photocounts
in their common signal beam. This result poses the question
whether the sub-Poissonianity of the marginal beams and the mutual
anti-correlations in photon-number fluctuations also occur in more
general, complex, TWB configurations. Here, we answer this
question by considering a straightforward generalization of the
above configuration in which three TWBs share their signal beams
for common photon detection and analyze the properties of the
three idler beams emerging after post-selection conditioned by
detecting a given number of common signal photocounts. We note
that, in our experiment, we do not exploit the entanglement in
polarization or spatial correlations between the beams. We only
rely on the correlations in the number of photons between the
signal and the idler beams. In our experiment, we divide the idler
beam into three parts in which the photon numbers are
independently measured. On the other hand the signal beam is
monitored as a whole and the detected number of signal photons is
used for post-selection.

The paper is organized as follows. Experimental setup is discussed
in Sec.~2. Reconstruction of the experimental histograms using
maximum-likelihood method and best fit by a multi-mode Gaussian
field is described in Sec.~3. In Sec.~3, also the quantities used
to characterize the post-selected fields including their
nonclassicality are introduced. Properties of the fields
post-selected by real detection are investigated in Sec.~4. The
properties of the fields arising in ideal post-selecting detection
are discussed in Sec.~5 using the experimental data and their
complete reconstruction. Conclusions are drawn in Sec.~6.

\section{Experimental setup}

\begin{figure}[t]
 \centerline{\includegraphics{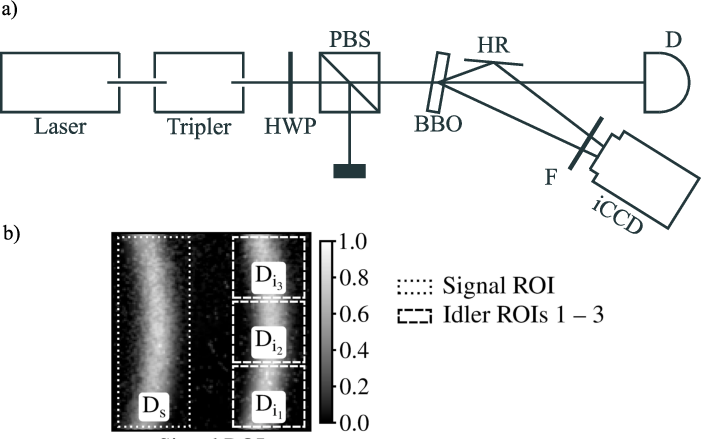}} \vspace*{8pt} \caption{(a)
 Schematic of the experimental setup composed of laser, tripler,
 half-wave-plate HWP, polarizing beam splitter PBS, BBO crystal,
 highly-reflected mirror HR, laser-light detector D, frequency
 filter F, and iCCD camera. (b) Multiple exposition of an
 image acquired by the detector - (left) signal strip monitored by
 detector D$_\mathrm{s}$, (right) idler strip divided into three
 parts detected by detectors D$_{\mathrm{i}_1}$,
 D$_{\mathrm{i}_2}$, and D$_{\mathrm{i}_3}$.}
\label{f:setup}
\end{figure}

The schematic of the experimental setup is shown in Fig.
\ref{f:setup}(a).  The light from the laser enters the frequency
tripler. The TWB is generated in type-I spontaneous parametric
down-conversion in a $\beta$-barium-borate crystal (BaB$_2$O$_4$,
BBO). Both beams of the TWB are detected by an iCCD camera as
shown in Fig. \ref{f:setup}(b). An interference bandpass filter is
placed in front of the camera. The signal and idler beams produce
stripes on the  photocathode of the iCCD. The area in which the
idler beam is detected is divided into three nearly equal parts.
This forms a triple TWB with common detection in the signal beam.

The femtosecond cavity-dumped Ti:sapphire laser with the central
wavelength of 840 nm serves as a light source. The wavelength at
the output of the frequency tripler is 280 nm. The laser system
repetition rate is 50 kHz and the pulse energy is 20 nJ. The iCCD
camera Andor DH345-18U-63 has the detection-window width of 7 ns
and is controlled by the synchronization pulses from the laser.
The camera's sampling frequency is approximately 14 Hz. The
bandpass interference filter F in front of the camera has the
spectral width of 14 nm at the central wavelength of 560 nm. The
pump intensity, and thus the TWB's intensity is actively
stabilized using the detector D, motorized half-wave plate HWP,
and polarization beam-splitter PBS.

Performing $1\times 10^6$ measurement repetitions we have arrived
at the experimental photocount histogram $f(
c_\mathrm{s},c_{\mathrm{i}_1}, c_{\mathrm{i}_2},
c_{\mathrm{i}_3})$ whose analysis gives comprehensive information
about the field properties.

\section{Reconstruction of the field}

In the experimental four-dimensional (4D) photocount histogram
$f(c_\mathrm{s}, c_{\mathrm{i}_1}, c_{\mathrm{i}_2},
c_{\mathrm{i}_3})$, the value $f( c_\mathrm{s}, c_{\mathrm{i}_1},
c_{\mathrm{i}_2}, c_{\mathrm{i}_3})$ gives the probability of
detecting $c_\mathrm{s}$, $c_{\mathrm{i}_1}$, $c_{\mathrm{i}_2}$,
and $c_{\mathrm{i}_3}$ photocounts in, in turn, the detection
areas D$_\mathrm{s}$, D$_{\mathrm{i}_1}$, D$_{\mathrm{i}_2}$, and
D$_{\mathrm{i}_3}$. In the reconstruction we reveal a 4D
photon-number distribution $p(n_\mathrm{s}, n_{\mathrm{i}_1},
n_{\mathrm{i}_2}, n_{\mathrm{i}_3})$ that describes the common
state of three multi-mode TWBs using the known detector
characteristics. The field reconstruction can symbolically be
illustrated as follows
\begin{equation}   
\label{eq:reconstruction}
 f(c_\mathrm{s},c_{\mathrm{i}_1}, c_{\mathrm{i}_2}, c_{\mathrm{i}_3}) \longrightarrow p( n_\mathrm{s}, n_{\mathrm{i}_1}, n_{\mathrm{i}_2}, n_{\mathrm{i}_3}).
\end{equation}
Modelling the process of photon detection we express the
photocount distribution $ f $ as a suitable linear function of
photon-number distribution $ p $ \cite{Perina1991} and this
relation has to be inverted using suitable approach, e.g. by
applying the maximum-likelihood (ML) approach.
We note that if the used detector were ideal, i.e., its detection
efficiency were 100$\%$ and it had no dark counts, the arrow in
Eq.~(\ref{eq:reconstruction}) would be replaced by an equal sign.

In addition to the reconstruction indicated in
Eq.~(\ref{eq:reconstruction}), it is possible to perform the
following partial reconstruction
\begin{equation}
\label{eq:preconstruction}
 f(c_{\mathrm{i}_1}, c_{\mathrm{i}_2}, c_{\mathrm{i}_3}; c_\mathrm{s}) \longrightarrow p(n_{\mathrm{i}_1}, n_{\mathrm{i}_2}, n_{\mathrm{i}_3}; c_\mathrm{s}).
\end{equation}
In this case, we fix the number $ c_\mathrm{s} $ of post-selected
signal photocounts and reconstruct the corresponding
three-dimensional (3D) photon-number distribution. Thus,
we reveal the 3D photon-number distributions of the optical fields
directly present in the experimental setup.

Analogs of these 3D photon-number distributions appropriate for an
ideal post-selecting detector are obtained from the reconstructed
4D photon-number distribution in Eq. (\ref{eq:reconstruction}).

We have applied two approaches to reconstruct the photon-number
distributions in Eqs. (\ref{eq:reconstruction}) and
(\ref{eq:preconstruction}). The first one is based upon a suitable
Gaussian fit of the three original TWBs.~\cite{PerinaJr2012a,
PerinaJr2013, PerinaJr2013a, PerinaJr2021} The second method
applies the ML approach.~\cite{PerinaJr2021}

In the Gaussian-fit reconstruction, we consider three TWBs
together and the photon-number distribution $ p_\mathrm{p} $ of their
common TWB composed of a paired and a noisy parts. The paired part
of the distribution is given as the convolution of three paired
parts belonging to individual TWBs:
\begin{equation}    
\label{eq:paired}
 p_\mathrm{p}( n_\mathrm{s}, n_{\mathrm{i}_1}, n_{\mathrm{i}_2}, n_{\mathrm{i}_3}) = 
\sum_{n_{{\rm s}_1}}^{n_{\rm s}}  \sum_{n_{{\rm s}_2}}^{n_{\rm s} - n_{{\rm s}_1}}
 p_{\mathrm{p}_1}(n_{\mathrm{s}_1}, n_{\mathrm{i}_1})  p_{\mathrm{p}_2}(n_{\mathrm{s}_2}, n_{\mathrm{i}_2})  p_{\mathrm{p}_3}(n_{\mathrm{s}} - n_{\mathrm{s}_1} - n_{\mathrm{s}_2}, n_{\mathrm{i}_3}),
\end{equation}
where $p_{\mathrm{p}_j}$ represent multi-mode thermal Mandel-Rice
distributions of photon pairs $p_{{\rm p}_j}(n_{{\rm s}_j}, n_{{\rm i}_j}) = \delta_{n_{\mathrm{s}_j}, n_{\mathrm{i}_j}}
p^{\mathrm{M-R}}(n_{\mathrm{s}_j};M_{\mathrm{p}_j}, B_{\mathrm{p}_j})$ for $\delta$ being the
Kronecker symbol and label $j$ indexing the TWBs, i.e. $ j= 1,2,3
$. The multi-mode thermal Mandel-Rice distribution takes the form
\begin{equation}    
\label{eq:MR}
 p^{\mathrm{M-R}}(n; M, B) = \frac{\Gamma(n + M)}{n! \Gamma(M)}
 \frac{B^n}{(1+B)^{n+M}},
\end{equation}
in which $M$ means the number of modes and $B$ denotes the mean
number of photons per mode.

To arrive at the Gaussian fit of the reconstructed photon-number
distribution, the paired part in Eq.~(\ref{eq:paired}) has to be
convolved with the noise part represented by the following four
noise distributions
\begin{eqnarray}   
 p(n_\mathrm{s}, n_{\mathrm{i}_1}, n_{\mathrm{i}_2}, n_{\mathrm{i}_3}) &=& \sum_{l_s=0}^{n_\mathrm{s}} p_{\mathrm{n}_\mathrm{s}}(n_\mathrm{s} - l_\mathrm{s})  \sum_{l_{\mathrm{i}_1}=0}^{n_{\mathrm{i}_1}} p_{\mathrm{n}_{\mathrm{i}_1}}(n_{\mathrm{i}_1} - l_{\mathrm{i}_1}) \sum_{l_{\mathrm{i}_2}=0}^{n_{\mathrm{i}_2}} p_{\mathrm{n}_{\mathrm{i}_2}}(n_{\mathrm{i}_2} - l_{\mathrm{i}_2}) \nonumber \\
 & & \sum_{l_{\mathrm{i}_3}=0}^{n_{\mathrm{i}_3}} p_{\mathrm{n}_{\mathrm{i}_3}}(n_{\mathrm{i}_3} - l_{\mathrm{i}_3}) p_\mathrm{p}(l_{\mathrm{i}_1}, l_{\mathrm{i}_2}, l_{\mathrm{i}_3}, l_\mathrm{s}).
\label{eq:recfield}
\end{eqnarray}
The noise photon-number distributions $p_{\mathrm{n}_{\mathrm{i}_j}}$ and $p_{\mathrm{n}_\mathrm{s}} $
for $j = 1, 2, 3$ introduced in Eq.~(\ref{eq:recfield}) are also
described by their corresponding Mandel-Rice distributions in
Eq.~(\ref{eq:MR}). All-together, the reconstructed photon-number
distribution $ p $ is described by 14 parameters introduced above:
$M_{\mathrm{p}_1}$, $M_{\mathrm{p}_2}$, $M_{\mathrm{p}_3}$, $M_{\mathrm{n}_\mathrm{s}}$, $M_{\mathrm{n}_{\mathrm{i}_1}}$,
$M_{\mathrm{n}_{\mathrm{i}_2}}$, $M_{\mathrm{n}_{\mathrm{i}_3}}$, $B_{\mathrm{p}_1}$, $B_{\mathrm{p}_2}$, $B_{\mathrm{p}_3}$,
$B_{\mathrm{n}_s}$, $B_{\mathrm{n}_{i_1}}$, $B_{\mathrm{n}_{i_2}}$, and $B_{\mathrm{n}_{i_3}}$. They
are determined \cite{PerinaJr2012a,PerinaJr2013a}, together with
the corresponding detection efficiencies, from the requirement of
equal first- and second-order intensity (photocount) moments taken
from the experiment and the Gaussian theoretical fit and minimal
declination between the experimental histogram and its theoretical
prediction.

Using the experimental data and detector parameters as summarized
in Table \ref{t:prm2}, we obtain the values of 14 parameters that characterize the
distribution $p( n_\mathrm{s}, n_{\mathrm{i}_1}, n_{\mathrm{i}_2},
n_{\mathrm{i}_3})$. These parameters are summarized in Table
\ref{t:prm}.

\begin{table}
 \caption{Detector properties.}
  \centering
  \begin{tabular}{lll}
    \toprule
    
    property &  D$_\mathrm{s}$
\hphantom{D$_{i_1}$} \hphantom{D$_{\mathrm{i}_1}$} & D$_{\mathrm{i}_1}$, D$_{\mathrm{i}_2}$,
D$_{\mathrm{i}_3}$\\
    \midrule
    number of active pixels  & 4536 & 1512 \\
detection efficiency & 0.233 & 0.226 \\
dark count rate & 0.220 & 0.073 \\

    \bottomrule
  \end{tabular}
  \label{t:prm2}
\end{table}

\begin{table}
 \caption{Parameters of the field reconstructed by the multi-mode Gaussian field.}
  \centering
  \begin{tabular}{llSSS}
    \toprule component
of field & subscript & $B$ & $M$ & $\langle n \rangle$ \\

    \midrule
		pairs s, i$_1$ & $p_1$ & 0.00475 & 552 & 2.62 \\
pairs s, i$_2$ & $p_2$ & 0.0910 & 29.9 & 2.71 \\
pairs s, i$_3$ & $p_3$ & 0.0524 & 51.5 & 2.70 \\
noise s & $n_s$ & 9.04 & 0.00779 & 0.070\\
noise i$_1$ & $n_{i_1}$ & 3.28 & 0.0274 & 0.089 \\
noise i$_2$ & $n_{i_2}$ & 402 & 6.33$\cdot 10^{-5}$ & 0.00025\\
noise i$_3$ & $n_{i_3}$ & 10.9 & 0.00225 & 0.024\\

    \bottomrule
  \end{tabular}
  \label{t:prm}
\end{table}

On the other hand, the ML reconstruction method
\cite{Dempster1977,Vardi1993} is iterative and starts with a
uniform distribution. After suitable number of steps the iteration
approaches its steady state that gives us the looked-for
photon-number distribution (for details, see \cite{PerinaJr2012}).

Then, the set of partially reconstructed distributions (for fixed
$ c_{\mathrm{s}} $) $p(n_{\mathrm{i}_1}, n_{\mathrm{i}_2}, n_{\mathrm{i}_3}; c_\mathrm{s})$, as described in
Eq.~(\ref{eq:preconstruction}), is derived along the formula
\begin{equation}   
 p(c_\mathrm{s}, n_{\mathrm{i}_1}, n_{\mathrm{i}_2}, n_{\mathrm{i}_3}) =  \sum_{n_\mathrm{s}=0}^{\infty} T(c_\mathrm{s}, n_\mathrm{s})
  p(n_\mathrm{s},n_{\mathrm{i}_1}, n_{\mathrm{i}_2}, n_{\mathrm{i}_3}),
\label{result}
\end{equation}
where the detection matrix $T(c_\mathrm{s}, n_\mathrm{s})$ gives
the probability that $c_\mathrm{s}$ photocounts are detected when
$n_\mathrm{s}$ photons arrive at the detector (for details, see,
e.g., Ref.~\cite{PerinaJr2012}). The elements of the detection
matrix $ T $ depend on the properties of the detector, namely the
number of active pixels, detection efficiency, and dark-count
rate.

The 3D photon-number distributions $ p_{c_\mathrm{s}}(n_{\mathrm{i}_1}, n_{\mathrm{i}_2},
n_{\mathrm{i}_3}) $ and $ p_{n_\mathrm{s}}(n_{\mathrm{i}_1}, n_{\mathrm{i}_2}, n_{\mathrm{i}_3}) $
characterizing the fields post-selected by detecting $ c_{\mathrm{s}} $
signal photocounts (real detection) and $ n_{\mathrm{s}} $ signal photons
(ideal detection) are derived from the above 4D photon-number
distributions by appropriate normalization:
\begin{eqnarray}   
 p_{c_\mathrm{s}}(n_{\mathrm{i}_1}, n_{\mathrm{i}_2}, n_{\mathrm{i}_3}) &=& p(c_\mathrm{s}, n_{\mathrm{i}_1}, n_{\mathrm{i}_2}, n_{\mathrm{i}_3})/
  \sum_{ n_{\mathrm{i}_1}, n_{\mathrm{i}_2}, n_{\mathrm{i}_3}} p(c_\mathrm{s},n_{\mathrm{i}_1}, n_{\mathrm{i}_2}, n_{\mathrm{i}_3}),
\label{eq:pcs}\\
 p_{n_\mathrm{s}}(n_{\mathrm{i}_1}, n_{\mathrm{i}_2}, n_{\mathrm{i}_3}) &=& p(n_\mathrm{s}, n_{\mathrm{i}_1}, n_{\mathrm{i}_2}, n_{\mathrm{i}_3})/
  \sum_{ n_{\mathrm{i}_1}, n_{\mathrm{i}_2}, n_{\mathrm{i}_3}}  p(n_\mathrm{s}, n_{\mathrm{i}_1}, n_{\mathrm{i}_2}, n_{\mathrm{i}_3}).
\label{eq:pns}
\end{eqnarray}
The corresponding 3D photocount histograms are determined as
\begin{equation}   
 f_{c_\mathrm{s}}(c_{\mathrm{i}_1}, c_{\mathrm{i}_2}, c_{\mathrm{i}_3}) = f(c_\mathrm{s},c_{\mathrm{i}_1}, c_{\mathrm{i}_2}, c_{\mathrm{i}_3})/
  \sum_{ c_{\mathrm{i}_1}, c_{\mathrm{i}_2}, c_{\mathrm{i}_3}} f(c_\mathrm{s}, c_{\mathrm{i}_1}, c_{\mathrm{i}_2}, c_{\mathrm{i}_3}).
\label{eq:pfs}
\end{equation}

\begin{figure}[t]   
 \centerline{\includegraphics[scale=0.7]{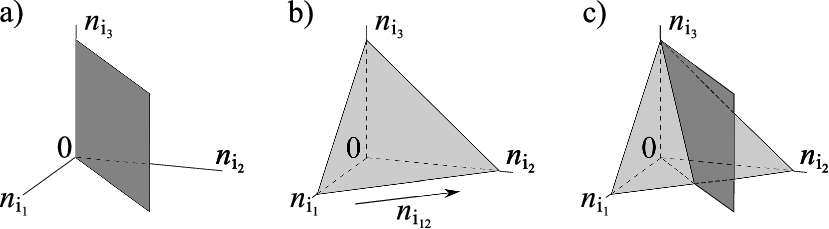}} \vspace*{8pt}
 \caption{Graphical representation of diagonal and triangular
  planes in the space ($n_{\mathrm{i}_1}$, $n_{\mathrm{i}_2}$, $n_{\mathrm{i}_3}$): (a)
  diagonal plane, (b) triangular plane, (c) both planes.}
\label{f:sections}
\end{figure}

We note that below we represent the main features of the
reconstructed post-selected 3D photon-number distributions
$p_{n_\mathrm{s}}$ and $p_{c_\mathrm{s}}$ using two representative planes (cuts) in
the whole 3D space spanned by the photon-numbers $n_{\mathrm{i}_1}$,
$n_{\mathrm{i}_2}$, and $n_{\mathrm{i}_3}$. These 'diagonal' and 'triangular' planes
are shown in Fig.~\ref{f:sections}. For clarity, coordinate $n_{\mathrm{i}_{12}}$ is used to describe triangular planes, the meaning of which is shown in Fig. \ref{f:sections}(b). 

Sub-Poissonianity of the post-selected beams is quantified using
the Fano factor
\begin{equation}   
 F_{n, \mathrm{i}_j} = \frac{\langle (\Delta n_{\mathrm{i}_j})^2\rangle}{\langle
 \Delta n_{\mathrm{i}_j} \rangle},
\label{eq:fano}
\end{equation}
where $ \langle (\Delta n)^2 \rangle \equiv \langle n^2\rangle -
\langle n\rangle^2 $ and $ \langle n^k\rangle $ gives a $k$-th
photon-number moment.

Anti-correlations between the photon-number fluctuations $ \Delta
n_{\mathrm{i}_j} $ and $ \Delta n_{\mathrm{i}_k} $ in the  $j$-th and $k$-th idler
beams are quantified by negative values of the following
correlation function
\begin{equation}  
 C_{\Delta n, \mathrm{i}_j\, i_k} = \frac{\langle{\Delta n_{\mathrm{i}_j}\,
  \Delta n_{\mathrm{i}_k}\rangle}}{\sqrt{\langle(\Delta n_{\mathrm{i}_j})^2\rangle \langle(\Delta n_{\mathrm{i}_k})^2\rangle}}.
\label{eq:core}
\end{equation}

Nonclassicality of the post-selected 3D idler fields is revealed
by suitable nonclassicality criteria (NCCa). The NCCa are
typically written as mathematical inequalities whose fulfillment
guarantees the field nonclassicality.~\cite{PerinaJr2022} The NCCa
inequalities are often written in terms of the moments of
integrated intensities.~\cite{PerinaJr2017a} There exist a large
number of various moment NCCa \cite{PerinaJr2017a, PerinaJr2020a}
derived by specific approaches and suitable for specific kinds of
fields. The nonclassicality of the fields built up from twin beams
is efficiently detected by the Cauchy-Schwarz and matrix
NCCa.~\cite{PerinaJr2022}

In our analysis, we apply the moment Cauchy-Schwarz NCC referred
to as $C_{111}^{000}$ and defined as
\begin{equation}   
 C_{111}^{000} = \langle W_{{\rm i}_1}^2 W_{{\rm i}_2}^2 W_{{\rm i}_3}^2\rangle  - \langle W_{{\rm i}_1} W_{{\rm i}_2} W_{{\rm i}_3} \rangle^2 < 0
\label{e:critmom}
\end{equation}
using the moments of intensities $W_{{\rm i}_1}$, $W_{{\rm i}_2}$,
and $W_{{\rm i}_3}$ associated in turn with the photon-numbers $
n_{\mathrm{i}_1} $, $ n_{\mathrm{i}_2} $, and $ n_{\mathrm{i}_3} $
and characterizing the field intensities at the detectors
D$_{\mathrm{i}_1}$, D$_{\mathrm{i}_2}$, and D$_{\mathrm{i}_3}$. In
parallel, we also use the intensity matrix NCC referred to as
$M_{101\, 010\,000}$ and given by the formula
\begin{eqnarray}  
 M_{101\, 010\,000} &=& \langle W_{{\rm i}_1}^2 W_{{\rm i}_3}^2\rangle \langle W_{{\rm i}_2}^2 \rangle
  + 2 \langle W_{{\rm i}_1} W_{{\rm i}_2} W_{{\rm i}_3}\rangle \langle W_{{\rm i}_2} \rangle \langle W_{{\rm i}_1} W_{{\rm i}_3} \rangle \nonumber \\
  &-& \langle W_{{\rm i}_1}  W_{{\rm i}_3}\rangle^2 \langle W_{{\rm i}_2}^2 \rangle -  \langle
  W_{{\rm i}_2}\rangle^2 \langle W_{{\rm i}_1}^2 W_{{\rm i}_3}^2 \rangle  - \langle W_{{\rm i}_1} W_{{\rm i}_2} W_{{\rm i}_3} \rangle^2 < 0.
\label{e:critmommat}
\end{eqnarray}

The moment NCCa can be transformed into their probability
counterparts using the Mandel detection
formula.~\cite{PerinaJr2022,PerinaJr2020a} The corresponding
inequalities are expressed in terms of probabilities of
photon-number distributions. As examples, we express the
probability NCCa $ \bar{C}_{111}^{000}$ and $ \bar{M}_{101\,
010\,000}$ corresponding to the intensity NCCa given in Eqs.
(\ref{e:critmom}) and (\ref{e:critmommat}):
\begin{eqnarray}  
 \bar{C}_{111}^{000} &=& 8 p_{c/n}(0,0,0) p_{c/n}(2,2,2)  - [ p_{c/n}(1,1,1)]^2 < 0,\\
 \bar{M}_{101\, 010\,000} &=& (2!)^3 p_{c/n}(2,0,2) p_{c/n}(0,2,0) p_{c/n}(0,0,0)\nonumber\\
  &+& 2 p_{c/n}(1,1,1) p_{c/n}(0,1,0) p_{c/n}(1,0,1) \nonumber \\
  &-& 2! [p_{c/n}(1,0,1)]^2 p_{c/n}(0,2,0)  - (2!)^2 [p_{c/n}(0,1,0)]^2 p_{c/n}(2,0,2) \nonumber \\
  &-& [p_{c/n}(1,1,1)]^2  p_{c/n}(0,0,0) < 0.
\label{e:critprob}
\end{eqnarray}
In Eq.~(\ref{e:critprob}), we substitute for $p_{c/n}$ either
$p_{c_\mathrm{s}}$ or $p_{n_\mathrm{s}}$ [Eqs.(\ref{eq:pcs}), (\ref{eq:pns})]
depending on whether the post-selection is performed according to
the number of photocounts or photons. General formulation of
probability NCCa and details can be found in
Refs.~\cite{PerinaJr2017a, PerinaJr2020a, PerinaJr2022}.

The NCCa not only identify the nonclassicality, they also provide
their quantification using the Lee nonclassicality depth
(NCD).~\cite{Lee1991} The NCD $\tau$ is derived from the
threshold value $s_{\rm th}$ of the ordering parameter at which the
corresponding quasi-distribution $ P $ of integrated intensities
begins to behave as a classical distribution observed by a given
NCC:
\begin{equation}   
\tau = (1 - s_{\mathrm{th}})/2.
\end{equation}
To determine the threshold value of $s_{\mathrm{th}}$ the moments of
integrated intensities as well as the photon-number distributions
have to be transformed to their general $ s $-ordered
forms.~\cite{Perina1991}

Using probability NCCa, we construct below 3D fields $\bar{\tau}
(n_{\mathrm{i}_1}, n_{\mathrm{i}_2}, n_{\mathrm{i}_3}) $ of NCDs. To determine the value of
NCD $\bar{\tau} $ at specific point $ (n_{\mathrm{i}_1}, n_{\mathrm{i}_2}, n_{\mathrm{i}_3}) $, we
consider only those NCCa that contain the probabilities in the
closest neighborhood of this point and take the greatest value of
the obtained NCDs. The 3D fields $\bar{\tau} (n_{\mathrm{i}_1}, n_{\mathrm{i}_2}, n_{\mathrm{i}_3})
$ of NCDs are then plotted in their diagonal and triangular
planes.

\section{Post-selection with real detector} \label{s:real}

We analyze the experimental 3D photocount histograms $f(
n_{\mathrm{i}_1}, n_{\mathrm{i}_2}, n_{\mathrm{i}_3};
c_\mathrm{s})$ considered as a function of the number $
c_{\mathrm{s}} $ of post-selecting signal photocounts and assign
to them the photon-number distribution
$p^\mathrm{ML}(n_{\mathrm{i}_1}, n_{\mathrm{i}_2},
n_{\mathrm{i}_3}; c_\mathrm{s})$ obtained by ML method and the
photon-number distribution $p^{\rm G}(n_{\mathrm{i}_1},
n_{\mathrm{i}_2}, n_{\mathrm{i}_3}; c_\mathrm{s})$ arising from
the multi-mode Gaussian fit.

The post-selected idler fields are characterized by their mean
photon numbers $\langle n_{{\rm i}_j}\rangle$ in an idler field,
their Fano factors $F_{n,\mathrm{i}_j}$, and their correlation
functions $C_{\Delta n, \mathrm{i}_j\, \mathrm{i}_k}$ between two
idler fields. We note that, in our fields, the mean numbers of the
idler fields, their Fano factors, and correlation functions behave
similarly in all three idler fields.

\begin{figure}[t]   
 \centerline{\includegraphics[scale=0.7]{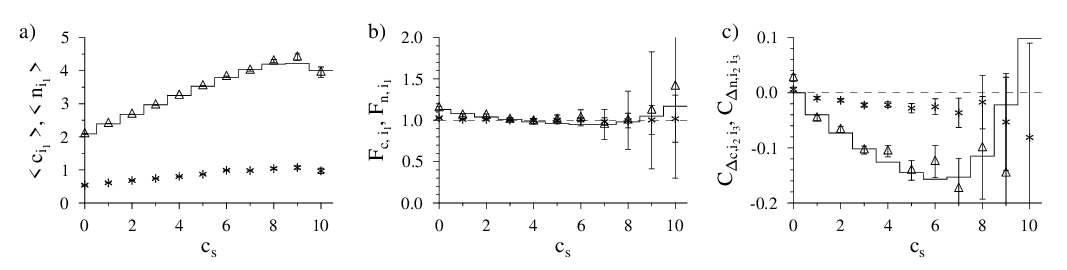}} \vspace*{8pt}
 \caption{(a) Mean photon (photocount) number $\langle n_{\mathrm{i}_1}
 \rangle $ ($\langle c_{\mathrm{i}_1} \rangle $) of idler beam 1, (b) its
 Fano factor $F_{n, \mathrm{i}_1}$ ($F_{c, \mathrm{i}_1}$), and (c) correlation
 function $ C_{\Delta n, \mathrm{i}_2\, {\rm i}_3} $ [$ C_{\Delta c, \mathrm{i}_2\, {\rm i}_3} $]
 of photon-number fluctuations in idler beams 2 and 3 as
 they depend on the post-selecting signal photocount number $c_\mathrm{s}$.
 Symbols $ \ast $ belong to the experimental photocount histogram,
 symbols $ \triangle $ to ML reconstruction, and solid curves to
 Gaussian reconstruction.}
\label{fa:mean}
\end{figure}

According to Fig.~\ref{fa:mean}(a), the mean photon numbers
$\langle n_{\mathrm{i}_1} \rangle $ increase with the increasing signal
photocount number $c_s$. On the other hand, the Fano factors
$F_{\mathrm{i}_1}$ decrease with the increasing $c_\mathrm{s}$ up to $c_\mathrm{s}=7$ and
then they increase, as shown in Fig.~\ref{fa:mean}(b). Similar
behavior is observed in Fig.~\ref{fa:mean}(c) for the correlation
function $C_{\Delta n, \mathrm{i}_2\, \mathrm{i}_3}$ that exhibits an increase of
anti-correlations between the idler beams 2 and 3 with the
increasing $c_\mathrm{s}$ up to $c_\mathrm{s}=7$ and then this anti-correlation
weakens. Such behavior reflects nonunit detection efficiency of
the post-selecting detector and its nonzero dark-count
rate.~\cite{PerinaJr2013b,PerinaJr2021} Whereas nonunit detection
efficiency has particularly negative influence at the Fano factor
$F_{n, \mathrm{i}_1}$ and correlation function $C_{\Delta n, \mathrm{i}_2\, \mathrm{i}_3}$ at
low post-selecting signal photocount numbers $c_\mathrm{s}$, nonzero
dark-count rate negatively affect the post-selection mechanism at
higher photocount numbers $c_\mathrm{s}$.

\begin{figure}[t]   
 \centerline{\includegraphics[scale=1.3]{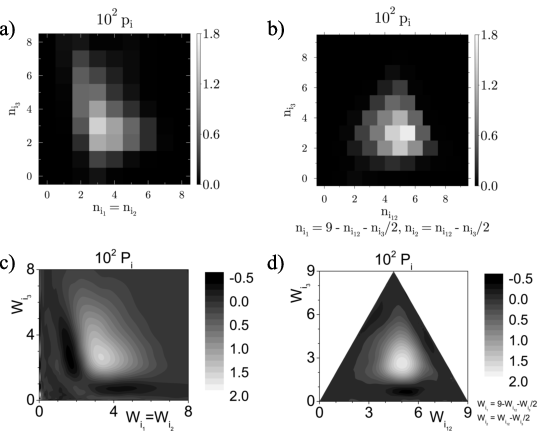}} \vspace*{8pt}
 \caption{Photon-number distribution $p_{\rm i}^\mathrm{ML}(n_{\mathrm{i}_1}, n_{\mathrm{i}_2},
 n_{\mathrm{i}_3})$ drawn in its (a) diagonal and (b) triangular planes and
 quasi-distribution $P_{\mathrm{i}}^\mathrm{ML}(W_{\mathrm{i}_1}, W_{\mathrm{i}_2}, W_{\mathrm{i}_3})$ of
 integrated intensities drawn in its (c) diagonal and (d)
 triangular planes for the field reached by real post-selection with $ c_\mathrm{s}=5 $. }
\label{fa:dis}
\end{figure}

\begin{figure}[t]   
 \centerline{\includegraphics[scale=0.35]{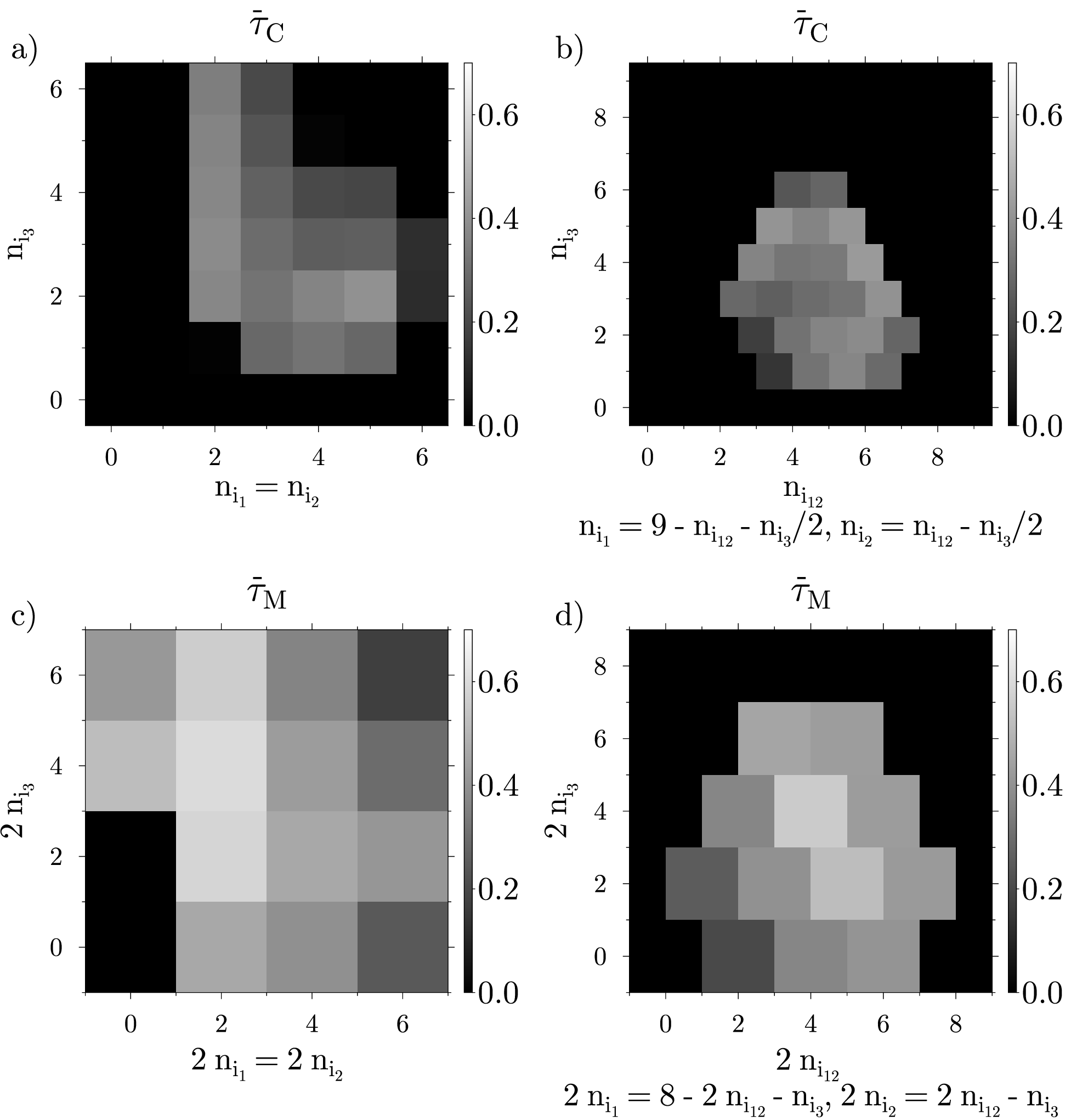}} \vspace*{8pt}
 \caption{Nonclassicality depth $ \bar{\tau}_\mathrm{C} $ [$ \bar{\tau}_\mathrm{M} $] for the probability Cauchy--Schwarz
  [matrix] NCCa drawn in its (a) [(c)] diagonal and (b) [(d)] triangular planes for the field reached by real post-selection
  assuming $ c_s=5 $.}
\label{fa:prob}
\end{figure}

To illustrate typical properties of the reconstructed fields, we
draw in Figs.~\ref{fa:dis}(a,b) the photon-number distribution
$p_{i}^\mathrm{ML}(n_{\mathrm{i}_1}, n_{\mathrm{i}_2},
n_{\mathrm{i}_3}) \equiv p_{c_{\rm s}
=5}^\mathrm{ML}(n_{\mathrm{i}_1}, n_{\mathrm{i}_2},
n_{\mathrm{i}_3})  $  of the field post-selected by detecting $
c_{\mathrm{s}} = 5 $ signal photocounts in its diagonal and
triangular planes. Nonclassicality of this field is certified
directly from the form of its quasi-distribution $
P_{\mathrm{i}}^\mathrm{ML}(W_{\mathrm{i}_1},W_{\mathrm{i}_2},W_{\mathrm{i}_3})
$ of intensities drawn in Figs.~\ref{fa:dis}(c,d) for the ordering
parameter $ s = 0.05$.~\cite{PerinaJr2021} In
Figs.~\ref{fa:dis}(c,d), regions with negative probabilities
located between the axes and the area with the maximal
probabilities are apparent. This quasi-distribution determined as
$ s = 0.05$ means that the true value of the NCD $ \tau $ is
greater than 0.475.
We compare this value of NCD $ \tau $ with those suggested by the
probability Cauchy--Schwarz NCCa and the matrix NCCa whose NCDs $
\bar{\tau} $ are drawn in Figs. \ref{fa:prob}(a---d) as they
depend on the location in the space of the idler-beam photon
numbers $ (n_{\mathrm{i}_1}, n_{\mathrm{i}_2},
 n_{\mathrm{i}_3}) $. Whereas the maximal NCD $ \bar{\tau}_C = 0.45 $ in the case of the
Cauchy--Schwarz NCCa, the matrix NCCa provides the greater maximal
NCD $\bar{\tau}_M = 0.56 $ certifying its better
performance.~\cite{PerinaJr2021}.
Better performance of the matrix NCCa compared to their
Cauchy--Schwarz counterparts is observed also when we apply the
corresponding intensity NCCa $C_{111}^{000}$ {($ \tau_{C} =
0.00$)} and $M_{101\, 010\,000}$ {($ \tau_{M} = 0.12 $)} given in
Eqs.~(\ref{e:critmom}) and (\ref{e:critmommat}). However, the
achieved values of NCDs $ \tau $ are considerably smaller than
when the probability NCCa are applied; the intensity
Cauchy--Schwarz NCC even does not indicate the nonclassicality.

\section{Post-selection with ideal detector}

Marginal sub-Poissonianity and anti-correlations of photon-number
fluctuations of the idler beams represent the prominent features
of the post-selected fields. Nevertheless, to observe them the
signal-beam detector used for post-selection has to have
sufficiently high detection efficiency. This is the reason why the
post-selected experimental idler fields did not exhibit their
sub-Poissonianity. To demonstrate the general features of these
states available by using sufficiently efficient detectors in
connection with the measured experimental histogram, we make
complete reconstruction of the 4D photon-number distribution
$p(n_{\rm s}, n_{{\rm i}_1}, n_{{\rm i}_2}, n_{{\rm i}_3})$ using
the measured histogram $f(c_{\rm s}, c_{{\rm i}_1}, c_{{\rm i}_2},
c_{{\rm i}_3})$. Then, fixing the number $ n_{\rm s} $ of signal
photons, we are left with the conditional photon-number
distribution $p_{n_{\rm s}}(n_{\mathrm{i}_1}, n_{\mathrm{i}_2},
n_{\mathrm{i}_3}) $. Such distribution corresponds to the
experimental situation in which an ideal detector is used for
post-selection. Similarly as above in Sec.~\ref{s:real}, we use
the iterative ML method and the Gaussian fit for reconstruction.

\begin{figure}[t]   
 \centerline{\includegraphics[scale=0.7]{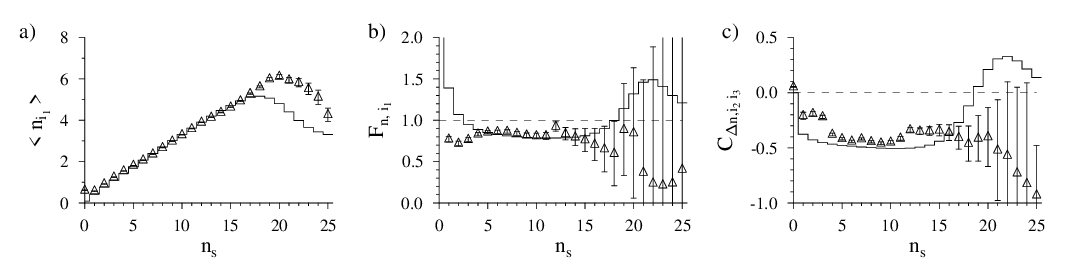}} \vspace*{8pt}
 \caption{(a) Mean photon number $\langle n_{\mathrm{i}_1}
 \rangle $ of idler beam 1, (b) its Fano factor $F_{n, \mathrm{i}_1}$, and (c)
 correlation function $ C_{\Delta n, \mathrm{i}_2\, {\rm i}_3} $
 of photon-number fluctuations in idler beams 2 and 3 as
 they depend on the post-selecting signal photon number $n_\mathrm{s}$ using ideal signal-beam detector.
 Symbols $ \triangle $ originate in ML reconstruction and solid curves
 arise in Gaussian reconstruction.}
\label{fb:mean}
\end{figure}

The mean number $\langle n_{{\rm i}_1}\rangle $ of photons in the
idler beam 1 increases approximately linearly with the number $
n_{\rm s} $ of post-selecting signal photons and we approximately
have $\langle n_{{\rm i}_1} \rangle \approx n_{\rm s}/3$ [see
Fig.~\ref{fb:mean}(a)]. Importantly, the Fano factors $F_{{\rm
i}_1}$ decrease with the increasing number $n_{\rm s}$ of
post-selecting signal photons up to $n_{\rm s} =13 $ and attain
their values lower than 0.8 [see Fig. \ref{fb:mean}(b)], which is
considerably lower than the quantum-classical border 1. Also the
correlation function $ C_{\Delta n, \mathrm{i}_2\, {\rm i}_3} $ of
photon-number fluctuations between the idler beams 2 and 3 attains
minimal negative values less than -0.45, indicating strong
anti-correlations [see Fig.~\ref{fb:mean}(c)].

\begin{figure}[t]   
 \centerline{\includegraphics[scale=1.3]{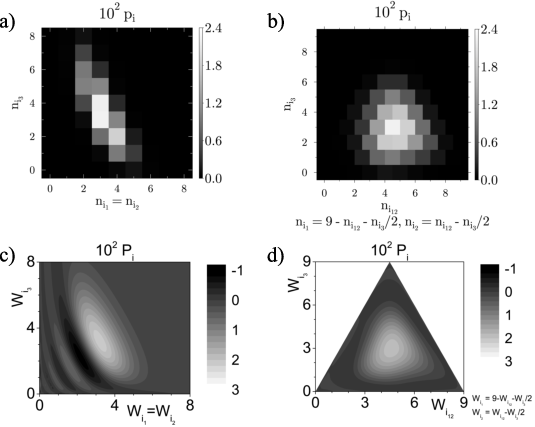}} \vspace*{8pt}
 \caption{Photon-number distribution $p_{\rm i}^\mathrm{ML}(n_{\mathrm{i}_1}, n_{\mathrm{i}_2},
 n_{\mathrm{i}_3})$ drawn in its (a) diagonal and (b) triangular planes and
 quasi-distribution $P_{\mathrm{i}}^\mathrm{ML}(W_{\mathrm{i}_1}, W_{\mathrm{i}_2}, W_{\mathrm{i}_3})$ of
 integrated intensities drawn in its (c) diagonal and (d)
 triangular planes for the idler field reached by ideal post-selection assuming $ n_\mathrm{s}=10 $. }
\label{fb:dis}
\end{figure}

\begin{figure}[t]   
 \centerline{\includegraphics[scale=0.35]{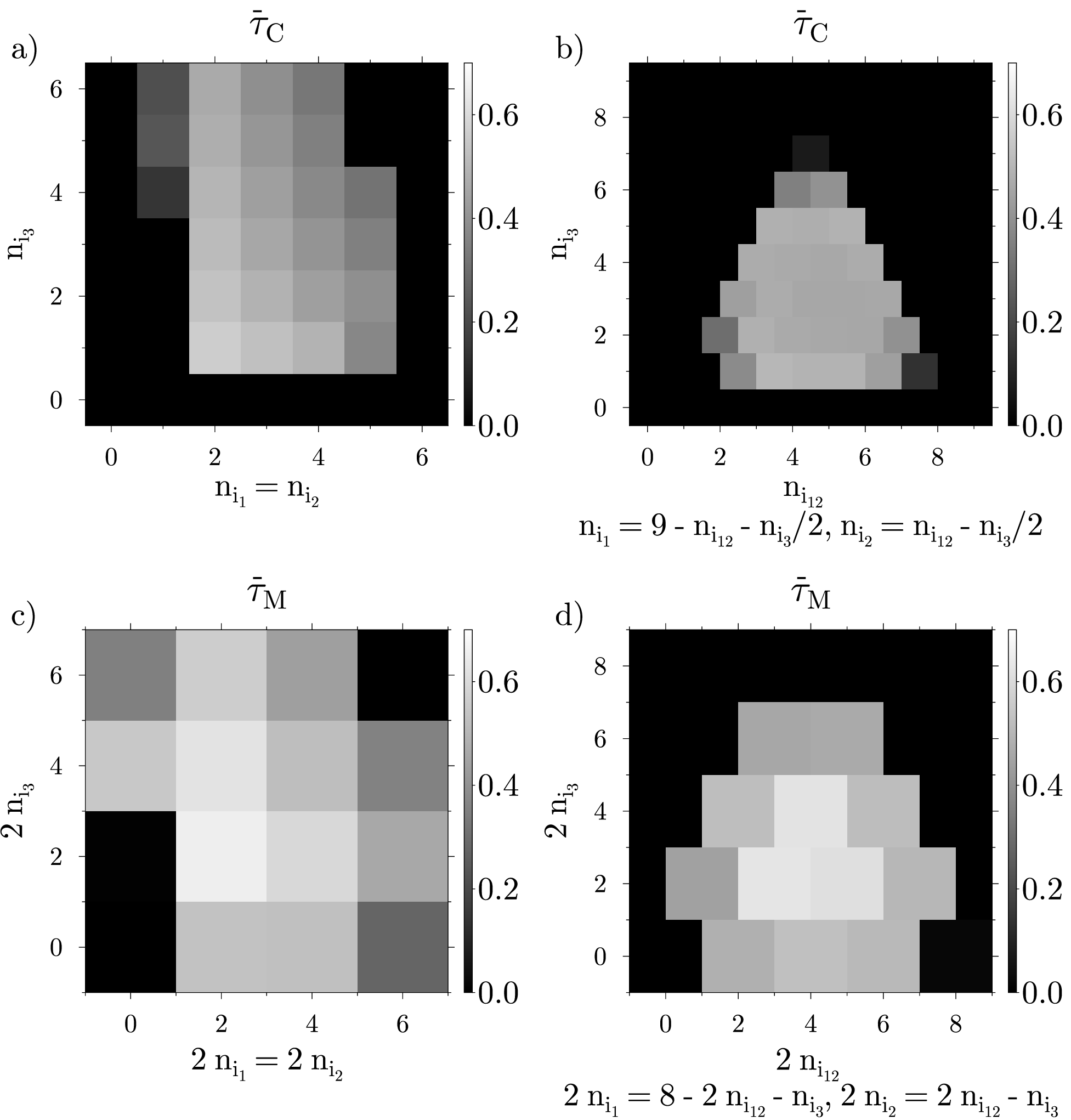}} \vspace*{8pt}
 \caption{Nonclassicality depth $ \bar{\tau}_\mathrm{C} $ [$ \bar{\tau}_\mathrm{M} $] for the probability Cauchy--Schwarz
  [matrix] NCCa drawn in its (a) [(c)] diagonal and (b) [(d)] triangular planes for the field reached by ideal post-selection
  assuming $ n_{\rm s}=10 $.}
\label{fb:prob}
\end{figure}

To demonstrate the properties of the analyzed fields in their most
general form, we plot in Figs.~\ref{fb:dis}(a,b) as an example the
photon-number distribution $p_{\rm i}^{\rm ML}(n_{\mathrm{i}_1},
n_{\mathrm{i}_2}, n_{\mathrm{i}_3}) \equiv p_{n_{\rm s} =10}^{\rm
ML}(n_{\mathrm{i}_1}, n_{\mathrm{i}_2}, n_{\mathrm{i}_3}) $ of the
field post-selected by detecting $ n_{\mathrm{s}} = 10 $ signal
photons in its diagonal and triangular planes in the form obtained
by the ML reconstruction. The corresponding quasi-distribution $
P_{\mathrm{i}}^\mathrm{ML}(W_{\mathrm{i}_1},W_{\mathrm{i}_2},W_{\mathrm{i}_3})
$ of integrated intensities calculated for $s = 0$ are shown in
Figs.~\ref{fb:dis}(c,d) in their diagonal and triangular planes
for comparison. Negative regions in these graphs mean that the
true value of NCD $ \tau $ is greater than 0.5. Comparing the
graphs of quasi-distributions $ P_{\mathrm{i}}^\mathrm{ML} $ drawn
in Figs.~\ref{fa:dis}(c,d) and \ref{fb:dis}(c,d) for the idler
fields reached by real and ideal post-selection, respectively,
quality of the reconstructed quasi-distribution $
P_{\mathrm{i}}^\mathrm{ML} $ obtained with ideal post-selecting
detector is much better. This also results in greater values of
the NCDs $ \tau $ belonging to both the probability and intensity
NCCa. The NCDs $ \bar{\tau}_{\rm C} $ obtained from the
Cauchy--Schwarz NCCa are drawn in Figs.~\ref{fb:prob}(a,b),
whereas the NCDs $ \bar{\tau}_{\rm M} $ of the matrix NCCa are
plotted in Figs.~\ref{fb:prob}(c,d). The reached maximal values of
NCDs $ \bar{\tau} $, $ \bar{\tau}_{\rm C} = 0.56 $ and $
\bar{\tau}_{\rm M} = 0.65 $, are greater than those reached by
real post-selection and belonging to the idler field post-selected
by $ c_{\rm s} = 5 $ signal photons analyzed in Sec.~
\ref{s:real}. Though the intensity Cauchy--Schwarz NCC
$C_{111}^{000}$ and matrix NCC $M_{101\, 010\,000}$ are less
efficient than their probability counterparts, they provide much
higher values of NCDs $ \tau $, {$ \tau_{\rm C} = 0.36 $ and $
\tau_{\rm M} = 0.38$} compared to the case analyzed in
Sec.~\ref{s:real}.

\section{Conclusions}

Using photon-number- and spatially- resolved detection of a twin
beam as provided by an iCCD camera we have obtained specific
three-beam states exhibiting mutual anti-correlations in
photon-number fluctuations of the beams and marginal
sub-Poissonian photon-number distributions. These states arise
after photon-number-resolved post-selection in a common signal
beam whose idler counterpart is divided into three spatially
separated idler beams of comparable intensity. Nonclassicality of
the post-selected fields is certified using suitable intensity
and probability nonclassicality criteria and their accompanying
nonclassicality depths as well as quasi-distributions of
integrated intensities. The Fano factor quantifies
sub-Poissonianity of the marginal beams. The properties of the
states obtained by using both real and ideal post-selecting
detectors are analyzed and mutually compared. 


Whereas the states obtained by ideal post-selection exhibit
both anti-correlations in photon-number fluctuations and marginal
sub-Poissonianity, lower detection efficiency of the real
post-selecting detector concealed the marginal sub-Poissonianity.
Nevertheless, the nonclassicality of the three-beam states
dominantly reflects the quantum correlations. These are strong
both when real and ideal post-selection is applied. Quantum
correlations are substantial for application potential of these
states.


The generated
states are potentially useful for metrology applications and well
as for different quantum-information protocols, e.g. in the area
of secure-quantum communications.

\section*{Acknowledgments}
The authors acknowledge support by the project OP JAC
CZ.02.01.01/00/22\_008/0004596 of the Ministry of Education,
Youth, and Sports of the Czech Republic.

\bibliographystyle{iopart-num}



\bibliography{sample}

\end{document}